\documentstyle[aps,prb]{revtex}
\input{psfig}
\begin{document}
\preprint{Version 26.8.1998}
\title{Phonon generation by current-carrying nanostructures}
\author{H. Totland$^{(1)}$, Y. M. Galperin$^{(1,2)}$, and
V. L. Gurevich$^{(2)}$}
\address{$^{(1)}$Department of Physics, University of Oslo,
                 P.~O.~Box 1048 Blindern, N--0316 Oslo, Norway \\
        $^{(2)}$Solid State Physics Department, A.~F.~Ioffe
Institute, 194021 Saint~Petersburg, Russia}
\maketitle

\begin{abstract}
We calculate the rate of acoustic phonon generation
by a current-carrying, ballistic quantum channel,
defined in a two-dimensional electron gas by a split-gate.
Both uniform and nonuniform channels are considered.
The generation rate of
acoustic phonons of a particular frequency and direction of propagation
is a steplike function of the applied bias voltage,
with threshold voltages that are calculated in the paper.
The emitted phonons have a characteristic angular distribution
which changes significantly at the thresholds.
The voltage dependence of the generation rate is shown to
be sensitive to the shape of the channel.
Thus the spectral and spatial distributions
of the emitted phonons bear information
both on electron-phonon coupling in the vicinity of the device
and on characteristics of the electron spectrum.
\end{abstract}


\section {Introduction}

Electronic properties of nanostructures 
can be effectively investigated by making use of
the interaction between electrons and phonons.
There has been done much experimental work on the response of
nanostructures to nonequilibrium ballistic phonons and to surface
acoustic waves,
see for example Refs.~\onlinecite{KNH,TSC} and references therein.
In this paper, however, we concentrate on the reverse effect,
namely on the emission of acoustic phonons from a nanostructure
carrying an electric current.
The purpose of the paper is to investigate
the spectral and spatial distribution of generated phonons
as a function of the voltage across a one-dimensional (or
quasi-one-dimensional) channel. The concrete results are obtained for
the case of a channel
defined in a two-dimensional electron gas (2DEG)
at the interface of a semiconductor heterostructure
by means of a split-gate at negative potential. Phonon emission from
various nanostructures has been extensively studied (see e.g.\
Refs.~\onlinecite{RDC,CWGS} and references therein). However, we are not
aware of any work on emission of acoustic phonons from a biased
quantum wire.

Using the Landauer-B\"uttiker-Imry approach,\cite{lby}
we consider a ballistic quantum channel
that connects two thermal reservoirs,
each of which being in an independent equilibrium state.
As was recently shown,\cite{gur1}
most of the heat from a current through the channel
is generated in the reservoirs.\cite{end0}
Nevertheless,
because of finite electron-phonon interaction inside the channel,
part of the heat should be generated by the nanostructure itself via
emission of phonons.
In the equilibrium there is a detailed balance
between the emitted and absorbed phonons.
However, our situation is nonequilibrium,
as the distributions of electrons
penetrating into a biased quantum channel from the leads
are characterized by different chemical potentials.
Therefore the phonon emission prevails over the absorption.
Such phonon emission results in nothing else than
the aforementioned generation of heat inside a ballistic nanostructure.

It is a complicated problem to calculate
the electron-phonon coupling constant for a realistic nanostructure
because of position-dependent screening
(see discussion in Sec.~\ref{secnonuniform}).
We do not attempt here to solve it
and assume the coupling constant to be known.

In Sec.~\ref{secuniform} we consider
phonons that are generated inside a long, uniform channel.
The effects of smooth edges are considered in Sec.~\ref{secnonuniform}.
As a result, one can describe the phonon emission
from realistic quantum channels,
which usually contain a long uniform part
with smoothly increasing width near the edges.

\section {Uniform channel} \label{secuniform}

In a uniform channel, the electron states can be represented as
$$\psi_{np} ({\bf r}) ={\cal L}^{-1/2}\chi_n ({\bf r}_\perp)\exp(i p
x/\hbar)\,,$$ where the normalization length ${\cal L}$ is assumed as
the channel length (i.e.\ the distance over which the potential
drops), $x$ and ${\bf r}_\perp$ are the longitudinal and
transverse directions respectively, $p$ is the $x$ component of the
electron quasimomentum, and $\chi_n ({\bf r}_\perp)$ is the wave
function of transverse quantization, the energy being $$\epsilon_n
(p) = \epsilon_n^0 +p^2/2m\, .$$ Here $m$ is the electron effective
mass, and $\epsilon_n^0\equiv\epsilon_n(p=0)$ is the bottom of the
$n$-th transverse band. To be concrete, we consider a channel
defined by a split-gate in a GaAs-AlGaAs heterostructure,
and therefore assume the system to be mechanically uniform
in the direction perpendicular to the interface.
Consequently, we consider bulk 3D acoustic phonons with
the displacement $\propto \exp(i {\bf qr})$. This assumption should be
modified in the case of acoustically nonuniform quantum wires where
confined modes can be present (see e.g.\ Ref.~\onlinecite{NAW}).
However, the main features of phonon generation remain similar.
The matrix element for phonon-induced transitions is defined as
$ C_{nn'}({\bf q}_{\perp}) =
\langle \chi_{n'}|\exp(-i{\bf q_{\perp}r_{\perp}})|\chi_n \rangle$,
where ${\bf q}$ is the phonon wave vector.

The number of phonons
 with wave vector $\bf q$, $N_{\bf q}({\bf r},t)$, is given by
the solution of the Boltzmann equation for phonons
(see e.g.\ Ref.~\onlinecite{gur}),
$$
  \frac{\partial N_{\bf q}}{\partial t}+{\bf s}\,\nabla N_{\bf q}
  ~=~{\cal R} \,.
$$
The phonon generation is described by the collision operator ${\cal R}
= \left(\partial N_{\bf q}/\partial t\right)_{\text{coll}}$, which can
be written as\cite{gur1}
\begin{eqnarray} \label{co2}
 {\cal R} &=& \nonumber
 \frac{2}{\cal A}\sum_{nn'} \int d\xi_p \, W_{\bf q}\,
 |C_{nn'}({\bf q}_{\perp})|^2
 \left[f_{n,p+\hbar q_x}(1-f_{n'p})(N_{\bf q} +1)
       - f_{n'p}(1-f_{n,p+ \hbar q_x})N_{\bf q}\right] \\
 && \quad\quad\quad\quad
 \mbox{}\times \delta
 [\epsilon_n(p + \hbar q_x)-\epsilon_{n'}(p) - \hbar \omega_{\bf q}] \, .
 \end{eqnarray}
Here ${\bf s}=\partial\omega_{\bf q}/\partial{\bf q}$ is the group
sound velocity, 
$d\xi_p = dp/2\pi \hbar$,
${\cal A}={\cal V}/{\cal L}$ is the cross section of the channel
(${\cal V}$ being its volume),
while the factor 2 comes from the summation over electron spin
(we assume all the transition probabilities to be spin-independent).
The coupling coefficient $W_{\bf q}$ for the piezoelectric coupling is
(cf.\ with Ref.~\onlinecite{GPI})
$W_{\bf q}=(\pi/\rho \omega_{\bf q})[e \beta_{q,lq} \nu_l ({\bf
q},a)/\epsilon_{qq}\epsilon_0]^2$.
Here $e$ is the electron charge, $\beta_{i,ln}$ is the tensor of
piezoelectric moduli, which is symmetric in the last two indices (see
e.g. Ref.~\onlinecite{gur}), $\epsilon_{il}$ is the tensor of
dielectric susceptibility, and $
{\bf \nu} ({\bf q},a)$ is the unit
polarization vector of the phonon branch $a$ with the wave vector $\bf q$.
The index $q$ indicates the projection of a tensor on the $\bf q$
direction, while $\rho$ is the mass density. For the deformation
potential interaction we have
$W_{\bf q}=\pi \Lambda^2q^2/\rho \omega_{\bf q}$, where $\Lambda$ is the
deformational potential constant for the phonon branch under consideration.
Using the approximate values for GaAs
($\beta=0.16\,{\text{Cm}^{-2}}$,
$\epsilon= 12$,
$\Lambda= 8\,{\text{eV}}$,
$s=\omega_{\bf q}/q= 3\times 10^5\,{\text{cm/s}}$),
we find that the piezoelectric interaction is the dominating one
for frequencies $\omega_{\bf q}\lesssim 5\times 10^{11}\,{\text{s}^{-1}}$.

Let us investigate the consequences of the energy and quasimomentum
conservation,
$$ \epsilon_n(p+\hbar q_x)-\epsilon_{n'}(p ) - \hbar \omega_{\bf q}=0\,.$$
For the solution of this equation, $p_{nn'}$, one has
\begin{equation} \label{p1i}
p_{nn'}=\frac{m}{\cos \theta}\left(s - \frac{
 \Delta_{nn'}}{\hbar q } \right) -\frac{\hbar q \cos \theta}{2} \,,
\end{equation}
where $s=\omega_{\bf q}/q$,
$\theta$ is the angle between $\bf q$ and the $x$-axis,
and $\Delta_{nn'} = \epsilon_n^0 - \epsilon_{n'}^0$.
Consequently, the delta-function in Eq.~(\ref{co2})
representing the energy and quasimomentum conservation can be expressed as
$(m/\hbar q |\cos \theta|)\, \delta (p-p_{nn'})$.

\begin{figure} 
\centerline{\psfig{figure=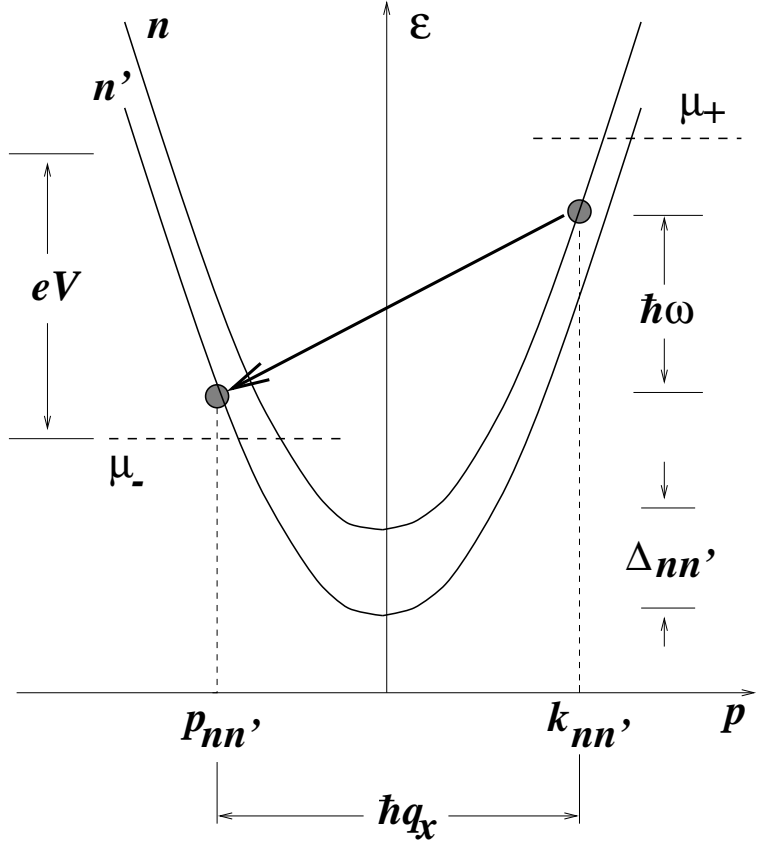,width=10cm}}
\vspace{0,5cm}
\caption{
Schematic representation of a transition
from an electron state of subband $n$
with energy $\epsilon$ and momentum $k_{nn'}$
to a state of subband $n'$
with energy $\epsilon-\hbar\omega$ and momentum $p_{nn'}=k_{nn'}-\hbar q_x$.
} 
\label{fig1}
\end{figure}

Following the Landauer-B\"uttiker-Imry approach,\cite{lby} we express
 the equilibrium distribution functions of the reservoirs as
$f^{(0)}_n (p) = f^{\text{(F)}}[\epsilon_n(p) - \mu^{(\pm)}]$, where
  $f^{\text{(F)}}$ is the Fermi function, $\mu^{(\pm)} = \mu \pm
 eV/2$, $\mu$ is the quasi Fermi level (depending on the gate
 voltage), while $V$ is the bias voltage.
Consider the transitions involving a phonon with a given $x$ component
  of the wave vector, $q_x >0$. Such a phonon can be emitted by
  a transition from an electron state having positive initial momentum
  $p+\hbar q_x$ to a state with negative momentum $p$
(see Fig.~\ref{fig1}). As a result, one gets\cite{gur1}
\begin{eqnarray*}
{\cal R} ~=~
\displaystyle{
\frac{m\, W_{\bf q}}{\pi{\cal A} \, \hbar^2 q|\cos \theta|}} \sum_{nn'}
|C_{nn'}({\bf q}_\perp)|^2
 \left(
f^{\text{(F)}}[\epsilon_n(k_{nn'})-\mu^{(+)}] \{1-f^{\text{(F)}}
[\epsilon_{n'}(p_{nn'})-\mu^{(-)}]\}(N_{\bf q} +1)\right.  && \\
\left.\mbox{} -
f^{\text{(F)}}[\epsilon_{n'}(p_{nn'})-\mu^{(-)}] \{1-f^{\text{(F)}}
[\epsilon_n(k_{nn'})-\mu^{(+)}]\}
N_{\bf q}\right) 
 \, ,
\quad &&
\end{eqnarray*}
where $k_{nn'}=p_{nn'}+\hbar q_x$.

\subsection{Intraband transitions}

Let us start by considering transitions within the same subband.
This is the dominating case for not too high phonon frequencies.
For $n=n'$, the solution $p_{nn'}$ of Eq.~(\ref{p1i})
is $n$-independent and equal to
\begin{equation} \label{p1}
p_1= ms/\cos \theta -(1/2)\, \hbar q \cos \theta\, .
\end{equation}
Thus the $n=n'$ part of the collision operator can be rewritten as
\begin{equation} \label{co3a}
  {\cal R} =
  \frac{mW_{\bf q}}{\pi {\cal A}\,\hbar^2 q \, |\cos \theta|} \, \sum_{n}
  |C_{nn}({\bf q}_{\perp})|^2
  \left[f_{n k_1}(1-f_{n p_1})(N_{\bf q} +1)-
  f_{n p_1}(1-f_{n k_1})N_{\bf q}\right] \,,
 \end{equation} 
where $k_1=p_1+hq_x=ms/\cos\theta+\hbar q_x/2$.

Let us consider the case $T=0$ (or, to be more specific, 
$\hbar\omega_{\bf q}\gg k_{\rm B}T$). Then Eq.~(\ref{co3a}) yields 
\begin{displaymath}
  {\cal R} =
  \frac{mW_{\bf q}}{\pi{\cal A}\, \hbar^2 q |\cos \theta |} \sum_{n} 
  |C_{nn}({\bf q}_\perp)|^2 \,
  \theta[\mu^{(+)}-\epsilon_n(k_1)] \, \theta[\epsilon_{n}(p_1)-\mu^{(-)}] \,.
\end{displaymath}
One can easily see that a {\em current-carrying channel can generate phonons}.
Since $p_1$ must be negative, Eq.~(\ref{p1}) leads to the inequality
\begin{equation} \label{cos1}
  \cos^2\theta \,>\, 2ms^2/\hbar\omega_{\bf q} \,,
\end{equation}
which defines an upper limit for the propagation angle.
Since $\epsilon_n (k_1)-\epsilon_n (p_1)=\hbar \omega_{\bf q}$,
$\mu^{(+)}-\mu^{(-)} =eV$, and $|\cos\theta|\le 1$,
the phonon generation is restricted to frequencies $\omega_{\bf q}$
satisfying the inequalities
\begin{equation} \label{2}
2ms^2 < \hbar\omega_{\bf q} < eV \,.
\end{equation}
The condition $\hbar\omega_{\bf q}<eV$ has been indicated 
earlier for the case of electro-phonon resonance in nanostructures.\cite{gur2}

The frequency region given by Eq.~(\ref{2}) is further restricted by
the Pauli principle via occupation numbers for the initial and final states.
The latter depend on the position of the quasi Fermi level $\mu$,
which in its turn is controlled by the gate voltage.
Let us consider intraband transitions within a given subband $n$.
The limitation is expressed by the inequalities
$ \epsilon_1+\hbar\omega_{\bf q}-eV/2
  < \mu <  \epsilon_1 +eV/2 $,
where $\epsilon_1 \equiv \epsilon_n[p_1(\omega_{\bf q},\theta)]$.
Consequently, if Eq.~(\ref{cos1}) is satisfied,
then phonons with frequency $\omega_{\bf q}$ and propagation direction $\theta$
are generated provided the bias voltage exceeds the threshold given by
\begin{equation} \label{Vthreshold}
  eV \,>\, 2\max(\mu-\epsilon_1,\,\epsilon_1+\hbar\omega_{\bf q}-\mu) \,.
\end{equation}

Let us investigate the frequency region in which
phonon generation is possible at least at some angle for a given $\mu$ and $V$.
If $\hbar\omega_{\bf q}>2ms^2$, then the function
$[p_1(\omega_{\bf q},\theta)]^2/2m$ varies between the value
$(\hbar\omega_{\bf q}-2ms^2)^2/8ms^2$ and zero
as the angle $\theta$ goes from zero to the upper limit from
Eq.~(\ref{cos1}).
We may thus summarize the above inequalities by stating that
phonon generation from transitions within the subband $n$
will take place (for at least some propagation angles)
in the frequency range
\begin{displaymath}
  2ms^2
  +\sqrt{8ms^2(\mu-eV/2-\epsilon_n^0)}\,\Theta(\mu-eV/2-\epsilon_n^0)
  \,<\, \hbar\omega_{\bf q} \,<\,
  \min(eV,\,\mu+eV/2-\epsilon_n^0) \,.
\end{displaymath}

\begin{figure}
\centerline{\psfig{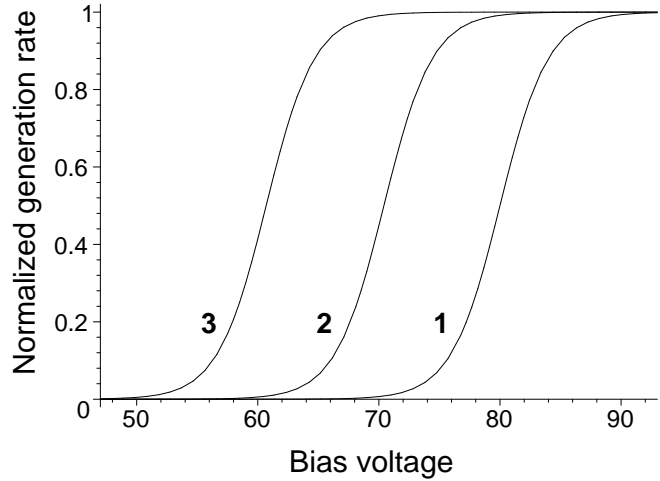}}
\vspace{-1cm}
\caption{
Voltage dependencies of the normalized phonon generation rate
for intraband transitions $n\rightarrow n$
and different propagation angles.
$\theta :\ 1 \ - \ 0^\circ, \ 2 \ - \ 30^\circ, \ 3 \ - \ 50^\circ$.
The bias voltage is measured in units of $k_BT/e$.
$\hbar\omega_{\bf q}/ms^2 = 6$,
$ms^2/k_BT=5$,
$\mu/\epsilon_n^0=1$.
}
\label{fig2}
\end{figure}

 In Fig.~\ref{fig2} the dependence of the phonon emission rate $\cal R$
 on the bias voltage is shown. Thresholds in the bias voltage are
 apparent.
If the voltage is further increased,
new subbands will contribute to the phonon emission
[see Eq.~(\ref{Vthreshold})].
Consequently, the generation rate is
a steplike function of the bias voltage.
There will be equally high steps corresponding to intraband transitions,
and a small modification from interband transitions.
The typical rate (corresponding to each step) is
${\cal R}=(ms/\hbar q|\cos\theta|){\cal R}_0$
with ${\cal R}_0=W_{\bf q}/\pi{\cal A} \hbar s$.
Assuming $\beta^2/\epsilon\epsilon_0\rho s^2=5\times 10^{-3}$,
$\hbar q/ms=6$, $s=3\times 10^5$ cm/s, $m =0.07m_0$,
${\cal A} = 10^{-12}$ cm$^2$, $\epsilon$=12,
we get ${\cal R}_0 \approx 10^{13}$ s$^{-1}$.
We assume that $qd\lesssim 1$, where $d$ is the width of the channel.
Thus $|C_{nn}({\bf q}_\perp)|^2\approx 1$, independently of the shape
 of the channel.

\begin{figure}
\centerline{\psfig{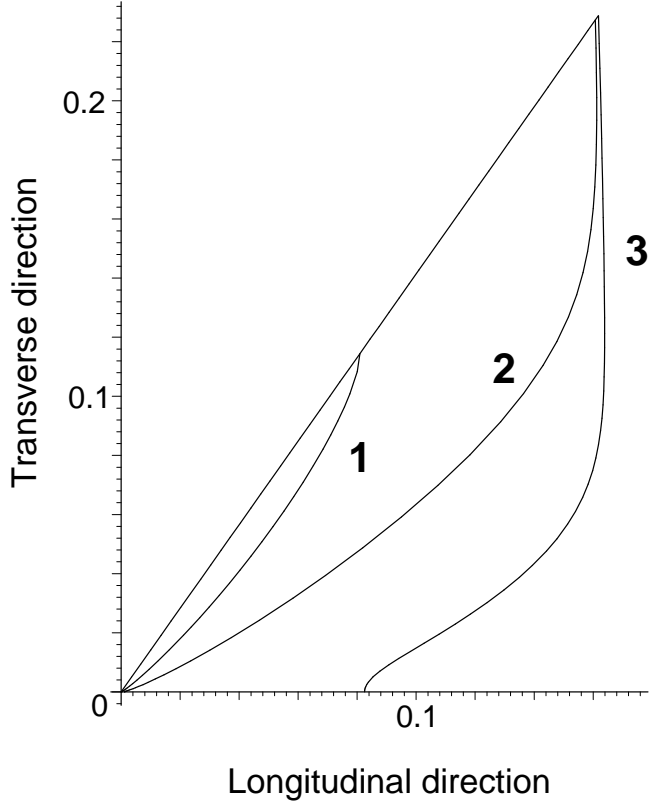}}
\vspace{0,5cm}
\caption{
Angular dependencies of the phonon generation rate
for intraband transitions $n\rightarrow n$
and voltages near the threshold.
$eV/k_BT: \ 1 \ - \ 60, \ 2 \ - \ 70, \ 3 \ - \ 80$.
The upper limiting angle corresponds to the condition
$p_1<0$, see Eq.~(\ref{cos1}).
The rate is measured in units of $W_{\bf q}/\pi{\cal A}\hbar s$.
Other values of the parameters are the same as in
\protect{Fig.~\ref{fig2}}.
We neglect
the angular dependence of the coupling parameter $W_{\bf q}$.
}
\label{fig3}
\end{figure}

Angular dependencies of the generation rate near the voltage threshold
are shown in Fig.~\ref{fig3}. It is seen that the character of the
angular dependence is changed at the threshold.
There is a sharp cutoff at the upper limiting angle
corresponding to $p_1(\omega_{\bf q},\theta)=0$, see Eq.~(\ref{cos1}).
This cutoff could be made more realistic (that is, smoother)
by taking into account
the scattering of electrons by a short-range impurity potential.
As was shown in Ref.~\onlinecite{BTG}, such a potential has the effect of
a smearing of the momentum conservation law.
Furthermore, we have neglected
the angular dependence of the coupling constant $W_{\bf q}$,
which depends on the acoustic properties of the substrate. This
additional dependence should be multiplied to obtain the observed
angular dependence.

\subsection {Interband transitions} \label{subsecinter}

The phonons can also be emitted by interband transitions.
Below we consider transitions from states
belonging to a subband $n$ to states of a different subband $n'$.
The frequency regions in which such processes may occur,
are different from those for intraband transitions,
although the inequality $\hbar\omega_{\bf q}<eV$ remains.
The possible propagation angles are further restricted by
the angular dependence of the factor $|C_{nn'}({\bf q}_\perp)|^2$.
Since we assume that $q_\perp d\lesssim 1$
($d$ being the channel width),
interband transitions are suppressed by this factor for all
directions. However, as will be shown, they bear an additional
information comparing to the intraband transitions.

The final electron momentum $p_{nn'}$ must be negative, as before,
but is now given by Eq.~(\ref{p1i}).
Furthermore, the initial momentum
$k_{nn'}=p_{nn'}+\hbar q\cos\theta$ must be positive.
This yields the condition
\begin{equation} \label{cos2}
  \cos^2\theta \,>\,
  (2ms^2/\hbar\omega_{\bf q})\,|1-\Delta_{nn'}/\hbar\omega_{\bf q}| \,,
\end{equation}
which replaces Eq.~(\ref{cos1}) for the upper limiting propagation angle.
In addition there is a voltage threshold given by Eq.~(\ref{Vthreshold}),
where $\epsilon_1$ now stands for
$\epsilon_{n'}[p_{nn'}(\omega_{\bf q},\theta)]$.

\begin{figure}
\centerline{\psfig{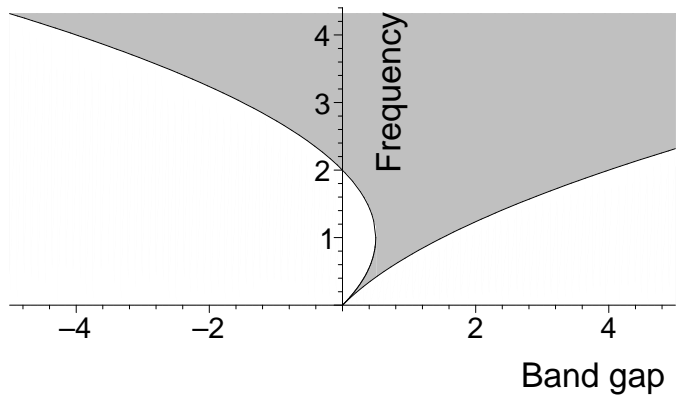}}
\vspace{-2,5cm}
\caption{
The shaded region corresponds to
the frequency region defined by Eqs.~(\ref{omega}).
There is no phonon generation for frequencies in the unshaded region.
The phonon frequency ($\omega_{\bf q}$)
is measured in units of $ms^2/\hbar$,
the band gap ($\Delta_{nn'}$) in units of $ms^2$.
}
\label{fig4}
\end{figure}

Phonon generation is only possible at frequencies for which
the right hand side of Eq.~(\ref{cos2}) is less than 1,
which occurs when
\begin{equation} \label{omega}
\begin{array}{lcl}
  \hbar\omega_{\bf q}>\epsilon_{++-} & \ \text{if} \ & \Delta_{nn'}<0 \,;\\
  \hbar\omega_{\bf q}>\epsilon_{++-} \ ~\text{or}~ \
  \epsilon_{+--}>\hbar\omega_{\bf q}>\epsilon_{-++}
  & \ \text{if} \ & 0<\Delta_{nn'}<ms^2/2 \,;\\
  \hbar\omega_{\bf q}>\epsilon_{-++} & \ \text{if} \ & ms^2/2<\Delta_{nn'} \,,
\end{array}
\end{equation}
with $ \epsilon_{\pm\pm\pm} = \pm ms^2 \pm
\sqrt{ms^2(ms^2\pm2\Delta_{nn'})}$.
This frequency region corresponds to the shaded area
in the $\omega_{\bf q}$--$\Delta_{nn'}$ diagram of Fig.~\ref{fig4}.
Since usually $|\Delta_{nn'}| \gg ms^2$,
phonon generation is restricted by the inequalities
\begin{displaymath}
\sqrt{2ms^2\,|\Delta_{nn'}|} \,\lesssim\, \hbar\omega_{\bf q} \,<\, eV \,.
\end{displaymath}

\begin{figure}
\centerline{\psfig{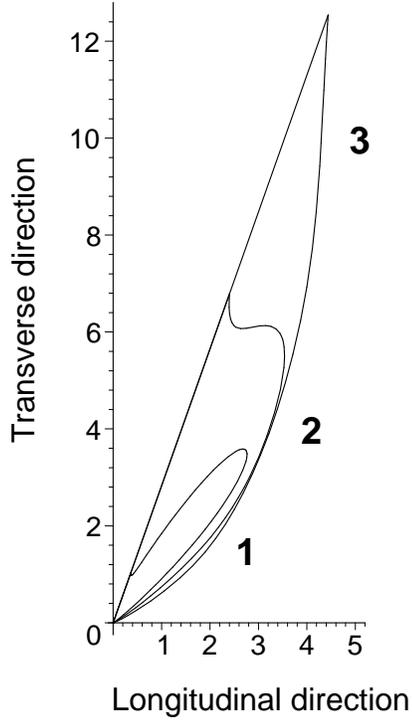}}
\vspace{1,6cm}
\caption{
Angular dependencies of the phonon generation rate in the interface plane
for interband transitions $n\rightarrow n'=n-1$ and different voltages.
$eV/k_BT: \ 1 \ - \ 35, \ 2 \ - \ 40, \ 3 \ - \ 45$.
The upper limiting angle corresponds to the condition
$k_{nn'}>0$, see Eq.~(\ref{cos2}).
The rate is measured in units of $(qd)^2W_{\bf q}/10^3\pi{\cal A}\hbar s$.
Furthermore, $\Delta_{nn'}/ms^2 = +8$, and
$\mu=\epsilon_{n}^0-\Delta_{nn'}/4$.
Other values of the parameters are the same as in
\protect{Fig.~\ref{fig3}}.
}
\label{fig5}
\end{figure}

In Fig.~\ref{fig5} one can see the angular dependencies of
the generation rate for three different values of the applied voltage.
They are not quite similar to the case of intraband transitions.
There is no phonon emission in the forward direction, because
the matrix element $C_{nn'}({\bf q}_\perp)$ vanishes for $\theta=0$.
The threshold values of voltage are shifted.
Furthermore,
the upper limiting angle is in this case due to the condition that
the initial momentum $k_{nn'}(\omega_{\bf q},\theta)$ must be positive.
The final momentum $p(\omega_{\bf q},\theta)$, on the other hand,
is negative for all angles, given these parameters.

\section {Nonuniform channel} \label{secnonuniform}

The edges of the channel play a specific role.
Namely, if the shape of the channel is smooth enough,
as we assume here,
one can use the so-called adiabatic approach,\cite{Les}
i.e.\ describe the situation in terms of
a position-dependent band structure.
We will show that
the phonons with a given frequency and propagation direction
can be generated only near specific points where
the local energy and quasimomentum conservation laws are met.
Consequently, the phonons emitted from the edges
bear information about the position-dependent band gaps
between the modes of transverse quantization.

Consider an adiabatic quantum channel with the width depending on 
coordinate $x$. The electron wave functions for such channels can be
subdivided into two categories -- the propagating states and the
reflected states on each side.
For the propagating state directed to the right ($p>0$), one has
\begin{displaymath}
  \psi_{\rightarrow}({\bf r})=
  \left|\frac{p}{p(x) {\cal L}}\right|^{1/2}\chi_{nx}({\bf r}_\perp)\,
  \exp \left[\frac{i} {\hbar} \int^x p(x') \, dx' \right]\,.
\end{displaymath}
In this expression, the transverse wave functions $\chi_n$
as well as the corresponding eigenvalues $\epsilon_n^0$
are assumed to vary slowly with the longitudinal coordinate $x$.
The longitudinal quasimomentum $p(x)\equiv p_n(x;\epsilon)$ is defined by
$p(x)^2/2m+\epsilon_n^0(x)=\epsilon>\max\{\epsilon_n^0(x)\}$,
while $p\equiv\sqrt{2m \epsilon}$ is the value at infinity
$x \rightarrow\pm\infty$.
Thus $\epsilon_n^0(x)\equiv\epsilon_n(p=0;x)$ is the local position of
the bottom of the $n$-th 1D transverse band at the point $x$.
We assume that the functions $\epsilon_n^0$ decrease
monotonically and symmetrically to zero
in both directions from the maximum point at $x=0$.
The oppositely directed propagating state $\psi_{\leftarrow}$
is defined in the same way, but with negative quasimomentum $p(x)$.

For a totally reflected electron state on the left hand side of the channel,
$\epsilon < \max \{\epsilon_n^0(x)\}=\epsilon_n^0(0)$,
the wave function is
\begin{displaymath}
  \psi_{\hookleftarrow}({\bf r})=
  \left|\frac{2 p}{p(x) {\cal L}}\right|^{1/2}\chi_{nx}({\bf r}_\perp)\,
  \sin \left[\frac{1}{ \hbar} \int^x_{x_t} p(x') \, dx'\right]\, , 
\end{displaymath}
$x_t$ being the classical turning point.\cite{end1}
A reflected state on the rhs of the channel is denoted by
$\psi_{\hookrightarrow}$.
Below we will characterize the electronic state by the combination
$\alpha = \{ n,p \}$ together with a subscript
$s=\rightarrow$, $\leftarrow$, $\hookleftarrow$, or $\hookrightarrow$.

As we are dealing with a nonuniform channel,
we must take into account the effect of screening.
The electrons in the wide regions of the channel
are more mobile than those in the narrow regions.
Therefore, the screening of the electron-phonon interaction,
and thus the coupling strength, will depend on the position.
We may take this effect into account by multiplying the function
$\exp(-i{\bf qr})$ in the transition amplitudes
by a dimensionless factor $\eta({\bf r})$.
This factor, which will be less than 1,
is determined by the screening of the
piezoelectric field or deformation potential field
by the electrons inside the leads and the channel.
In the 2D leads, $\eta({\bf r})$ is small
because of the high conductance of the 2DEG.
However, if the channel width is of the order of
the effective Bohr radius $\epsilon\epsilon_0 \hbar^2/me^2$,
then the screening inside the channel is not too strong.
The screening of the effective field in realistic
gated structures is far from being satisfactory understood, 
the more so that we are actually dealing with a nonstationary effect.
However, there are indications that
the effective potential is not much lower than the unscreened
one (see e.g.\ Ref.~\onlinecite{wix}), and we shall assume this
throughout the paper.
In a narrow channel and at $q{\cal L} \gg 1$ (here we denote by $\cal L$
the effective length of the channel),
 one can consider
$\eta({\bf r})$ as a smooth function $\eta(x)$ inside the channel
and rapidly decreasing outside.

Another simplification that arises from the above inequality is that
one can employ the {\em stationary phase approximation} for
estimation of the transition probabilities.
For two propagating states
$\alpha_s=\{n,p_\alpha\}_s$ and $\beta_{s'}=\{n',p_\beta\}_{s'}$,
the transition amplitude is
\begin{equation}  \label{int}
 \langle \beta_{s'}| \eta(x)\exp(-i{\bf qr}) | \alpha_s \rangle
  = \int_{-\infty}^{\infty} dx\, A(x) \exp[i \varphi(x)]\,,
\end{equation}
where $A(x)$ is a result of the integration in the transverse directions,
while the $x$-dependent part of the integrand's phase is given by
\[
  \varphi(x) \equiv \varphi_{\alpha s,\beta s'}(x) =
  \hbar^{-1}\int^x\left[p_\alpha(x')-p_\beta (x')\right]\,dx'-q_x x\,.
 \]
Here $p_\alpha(x)$ is determined as the solution $p$ of the equation
$\epsilon_n(p,x)=p_\alpha^2/2m$ $(=\epsilon_\alpha)$.
With one or both of the states being reflected ones, we get
two or four terms in the transition amplitude, respectively.
Now the phase $\varphi(x)$ is expanded
around a stationary point $x^*$ defined by the equation $d \varphi/dx
=0$, 
\begin{equation}  \label{varphiexp}
\varphi(x)=
\varphi(x^*)+\varphi''(x^*)(x-x^*)^2/2\, .
\end{equation}
If such a stationary point $x^*$ exists for $\varphi$, then
the main contribution to the integral is concentrated around this point.
As the part $A(x)$ of the integrand is assumed to vary slowly
on the scale $\sim\sqrt{2/|\varphi''(x^*)|}$,
one can substitute $A(x)$ in the integrand by $A(x^*)$.
If $\varphi$ has no stationary points,
one assumes a rapidly oscillating phase everywhere,
thus the total contribution to the transition amplitude is much
smaller than above.

In this picture, the transitions are localized at the points $x^*$
where $\varphi'(x^*)=0$. In our case, the change in energy
$\epsilon_\alpha=\epsilon\rightarrow
\epsilon_\beta=\epsilon-\hbar\omega_{\bf q}$
due to emission of a phonon is accompanied by
a quasimomentum transfer $-\hbar q_x$, i.e.\ at the point of stationary
phase one arrives at the {\em local conservation condition}
\begin{equation} \label{cl1}
p_n(x^*;\epsilon)=p_{n'}(x^*;\epsilon-\hbar\omega_{\bf q})+\hbar q_x \,.
\end{equation}
We would like to emphasize once again that the {\em local} values of
quasimomentum entering Eq.~(\ref{cl1})---rather than the asymptotic
values $p$---determine the conservation law.

The above approach has been employed to analyze
the photoconductance\cite{GGKS} and the acoustoconductance\cite{TBG}
in an adiabatic quantum channel.

\subsection{Phonon generation}

The transition amplitude of Eq.~(\ref{int}) is given by
\begin{displaymath}
  \langle \beta_{s'}| \eta(x)e^{-i{\bf qr}} | \alpha_s \rangle \,=\,
  \int_{-\infty}^\infty dx \, C_{nn'}({\bf q}_\perp,x) \,
  \tilde{\eta}_{\alpha s,\beta s'}(x) \,
  \exp[i\varphi_{\alpha s,\beta s'}(x)] \,,
\end{displaymath}
where $C_{nn'}$ depends (smoothly) on $x$ through the transverse functions
$\chi_{nx}$, while
\begin{displaymath}
  \tilde{\eta}_{\alpha s,\beta s'}(x) \,\equiv\, \eta(x) \, \xi_s
  \xi_{s'} 
  \left|p_\alpha p_\beta/p_\alpha(x)p_\beta(x){\cal L}^2\right|^{1/2}.
\end{displaymath}
The factor $\xi_s$ is 1 for propagating states,
$\pm i/\sqrt{2}$ for reflected ones.
In the stationary phase approximation, the functions
$C_{nn'}({\bf q}_\perp,x)$ and $\tilde{\eta}_{\alpha s,\beta s'}(x)$
are replaced by their values at the stationary point
$x^*_{\alpha s,\beta s'}$. Writing
\begin{displaymath}
  \Phi_{\alpha s,\beta s'} \equiv
  \int_{-\infty}^\infty\exp[i\varphi_{\alpha s,\beta s'}(x)] \, dx \,,
\end{displaymath}
we get for the generation rate of the nonuniform channel
\begin{eqnarray} \label{co3b}
  {\cal R} &=&
  (2W_{\bf q}/{\cal V}) \, \sum_{\alpha s,\beta s'}
  |\Phi_{\alpha s,\beta s'}|^2 \,
  |\tilde{\eta}_{\alpha s,\beta s'}(x^*_{\alpha s,\beta s'})|^2 \,
  |C_{nn'}({\bf q}_{\perp},x^*_{\alpha s,\beta s'})|^2  \nonumber\\
  && \quad\quad\quad \mbox{}\times \left[ f_{\alpha s}(1-f_{\beta s'})
  (N_{\bf q} +1) - f_{\beta s'}(1-f_{\alpha s})N_{\bf q} \right] \,
  \delta(\epsilon_\alpha-\epsilon_\beta-\hbar\omega_{\bf q}) \,,
\end{eqnarray}
where ${\cal V}$ is the interaction volume
(which is not known precisely,
but is of the order of the geometrical volume of the channel).
The point $x^*$ and the phase $\varphi(x)$ are dependent on
$\alpha_s$ and $\beta_{s'}$. However,
to avoid excessive proliferation of indices we will often omit these indices.

Now assuming that the second order term
in the expansion (\ref{varphiexp}) of the phase
grows faster to values of the order 1 than the following terms
(i.e.\ at $|\varphi'''(x^*)/\varphi''(x^*)^{3/2}|\ll 1$),
the factor $\Phi_{\alpha s,\beta s'}$ can be evaluated as
\[
  \Phi_{\alpha s,\beta s'} ~=~
  \left|2\pi/\varphi''(x^*)\right|^{1/2}
  \exp \left[ i\varphi(x^*)+i(\pi/4)\, \text{sgn}
  \, \varphi''(x^*)\right] \,,
\] 
where $\text{sgn}\, x = x/|x|$.
The quasimomenta $p_\alpha(x^*)$ and $p_\beta(x^*)$ at the transition point
are fixed by the conditions
\begin{displaymath}
  \epsilon_n^0(x^*) + \frac{p_\alpha(x^*)^2}{2m} \,=\,
  \epsilon_{n'}^0(x^*) + \frac{p_\beta(x^*)^2}{2m} + \hbar\omega_{\bf q}
\,,\quad  p_\alpha(x^*) \,=\, p_\beta(x^*) + \hbar q_x  \,.
\end{displaymath}
They can be positive or negative.
In case of one or both of the involved states being reflected ones,
the transition amplitude falls into two or four parts, respectively,
since a reflected state is a superposition of waves with $p>0$ and $p<0$.
However, if $n=n'$,
a stationary point $x^*$ exists for at most one of these terms, and
we denote by $+p_{\alpha(\beta)}$ the momentum with the corresponding sign.

\subsection{Intraband transitions}

For $n=n'$ we have
\begin{equation} \label{palpha_pbeta}
  p_\alpha(x^*)=k_1\equiv ms/\cos\theta+\hbar q_x/2 \,,\quad
  p_\beta(x^*) =p_1\equiv ms/\cos\theta-\hbar q_x/2 \,.
\end{equation}
The quasimomenta at the transition point are the same as in the case of a
uniform channel.
However, the electron energy $\epsilon_\alpha$
is no longer fixed by the phonon $(\omega,{\bf q})$,
since there are now different solutions $x^*$ of Eqs.~(\ref{palpha_pbeta})
corresponding to different energies
$\epsilon_\alpha=\epsilon_n^0(x^*)+k_1^2/2m$.

\begin{figure}
\centerline{\psfig{figure=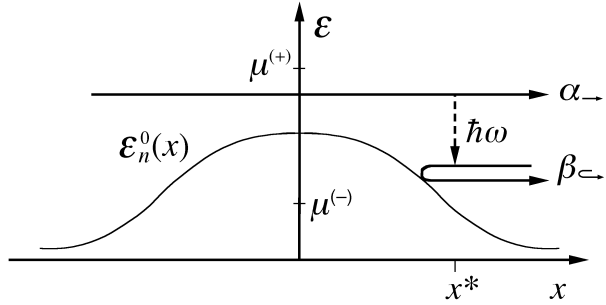,width=8cm}}
\vspace{0,2cm}
\caption{
Phonon emission in a nonuniform channel
by a transition from a propagating to a reflected electron state.
The transition is localized around the point $x^*$.
}
\label{fig6}
\end{figure}

\begin{figure}
\centerline{\psfig{figure=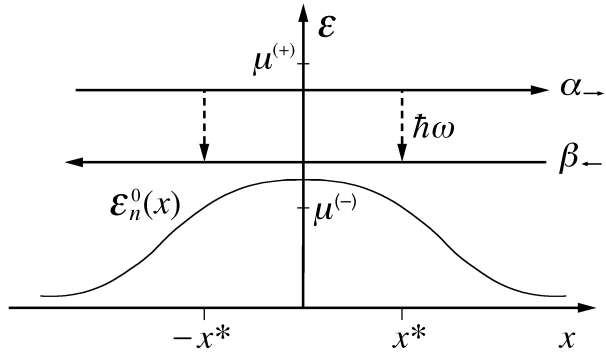,width=8cm}}
\vspace{0,2cm}
\caption{
Phonon emission in a nonuniform channel
by a transition between two oppositely directed propagating states.
There are two transition points.
}
\label{fig7}
\end{figure}

Let us again assume $T=0$. Then the only possible phonon-generating
processes are the ones shown in Figs.\ \ref{fig6} and \ref{fig7}.
Thus in Eq.~(\ref{co3b}) we have $s=\rightarrow$,
while $s'=\leftarrow$ (if $\epsilon_\alpha-\hbar\omega_{\bf q}>\epsilon_n^0(0)$)
or $s'=\hookrightarrow$ (if $\epsilon_\alpha-\hbar\omega_{\bf q}<\epsilon_n^0(0)$).
Introducing the density of states $g(\epsilon)$, we get
\begin{eqnarray}
  {\cal R} &=&
  (2W_{\bf q}/{\cal V}) \, \sum_n \int d\epsilon\,
  g(\epsilon)\,g(\epsilon-\hbar\omega_{\bf q})\,|\Phi_{n\epsilon}|^2\,
  |\tilde{\eta}_{n\epsilon}(x^*_{n\epsilon})|^2\,
  |C_{nn}({\bf q}_\perp,x^*_{n\epsilon})|^2 \nonumber\\
  && \quad\quad\quad\quad\quad\quad\quad
  \mbox{}\times \Theta(\mu^{(+)}-\epsilon)\,
  \Theta(\epsilon-\hbar\omega_{\bf q}-\mu^{(-)})\,, \label{conuc1}
\end{eqnarray}
where the following three restrictions are implied:
(i) A transition point $x^*$ must really exist for the energy $\epsilon$
involved, which is only the case when
$k_1^2/2m<\epsilon<k_1^2/2m+\epsilon_n^0(0)$.
(ii) The initial state must be propagating, thus $\epsilon_n^0(0)<\epsilon$.
(iii) For a transition between propagating states (Fig.~\ref{fig7}),
the final one must be propagating to the left.
This means that $p_1<0$, which, as before, leads to Eq.~(\ref{cos1}),
and hence the condition $\hbar\omega_{\bf q}>2ms^2$,
for such transitions.
However, for transitions to a nonpropagating state (Fig.~\ref{fig6}),
there is no such restriction.

For a transition of the kind in Fig.~\ref{fig7}, there will in fact be
{\em two} transition points, one on each side of the constriction
(cf.\ with Ref.~\onlinecite{MG}).
The corresponding two parts of the transition amplitude
give rise to the interference term
$2\{1+\sin[\varphi(x^*)-\varphi(-x^*)]\}$.

If, as in the case of a parabolic or square confining potential,
the transverse energy has the form $\epsilon_n^0(x)=\alpha_n/d(x)^2$,
$d(x)$ being the width of the channel, then
\[
  \varphi''(x^*) = \frac{p_\alpha'(x^*)-p_\beta'(x^*)}{\hbar}
  = \frac{2mq_x\,\epsilon_n^0(x^*)}{p_1 k_1} \frac{d'(x^*)}{d(x^*)}\,.
\]
Inserting these expressions and
$g(\epsilon)={\cal L}\sqrt{m}/h\sqrt{2\epsilon}$ into Eq.~(\ref{conuc1}),
we finally get
\begin{equation} \label{conuc2}
  {\cal R} = 
  \frac{mW_{\bf q}}{\pi{\cal V}\hbar^2q|\cos\theta|} \sum_n \int d\epsilon \,
  \frac{\eta^2(x^*_{n\epsilon}) \, |C_{nn}(q_\perp,x^*_{n\epsilon})|^2 \,
  d(x^*_{n\epsilon})}
  {\epsilon_n^0(x^*_{n\epsilon}) \, |d'(x^*_{n\epsilon})|}
  \, \nu_n(\epsilon) \,,
\end{equation}
where the integration goes from
$\max(k_1^2/2m,\,\epsilon_n^0(0),\,\mu^{(-)}+\hbar\omega_{\bf q})$ to
$\min(k_1^2/2m+\epsilon_n^0(0),\,\mu^{(+)})$,
and $\nu_n(\epsilon)$ represents the interference part.
This function is given by
\[
  \nu_n(\epsilon) = \left\{1+\sin\left[2\hbar^{-1}\int_0^{x^*}dx
  \left(|p_n(x;\epsilon)|+|p_n(x;\epsilon-\hbar\omega_{\bf q})|-\hbar q_x
  \right)\right]\right\} \Theta(\hbar\omega_{\bf q}\cos^2\theta-2ms^2)
\]
if $\epsilon>\epsilon_n^0(0)+\hbar\omega_{\bf q}$,
otherwise $\nu_n(\epsilon)=1/4$.
As before, the Fermi functions
$f^{\text{(F)}}(\varepsilon)\rightarrow\Theta(-\varepsilon)$
in Eq.~(\ref{conuc1})
cause the condition $\hbar\omega_{\bf q}<eV$.

When the energy $\epsilon$ approaches $k_1^2/2m$ (from above)
or $\epsilon_n^0(0)+k_1^2/2m$ (from below),
the transition point $x^*_{n\epsilon}$
approaches infinity or zero, respectively.
At these energies, which correspond to regions where
the stationary phase approximation is not valid,
$\epsilon_n^{0}(x^*)$ and $d'(x^*)$, respectively, approach zero,
making the integrand in Eq.~(\ref{conuc2}) diverge.
However, these energy regions make up only two narrow parts
of the integration range, and the integral itself converges.

\begin{figure}
\centerline{\psfig{figure=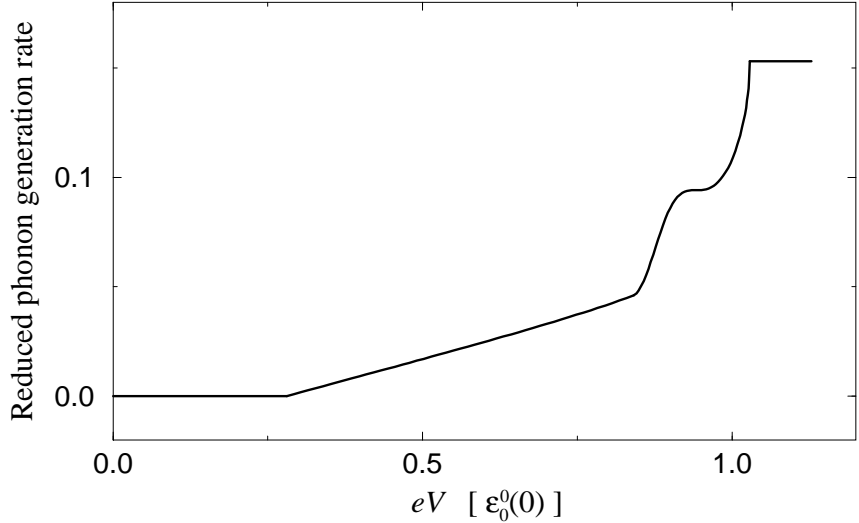,width=12cm}}
\vspace{0,4cm}
\caption{
Dependence of the phonon generation rate
$\cal R$ in a nonuniform channel
on the bias voltage $V$,
corresponding to intraband transitions of the zeroth subband ($n=0$).
Phonon generation sets on when $eV$ exceeds $\hbar\omega_{\bf q}$
and increases approximately linearly,
corresponding to transitions like the one in Fig.~\ref{fig6}.
In regions where transitions between propagating states
are possible as well (Fig.~\ref{fig7}),
the generation rate increases nonlinearly,
corresponding to the interference connected with these processes.
When $\mu^{(+)}=\mu+eV/2$ becomes large enough,
no further transitions are possible,
and the generation rate stays constant
until the onset of contributions from the next subband (not shown).
Here, the channel width is assumed to vary as $d(x)=d(1+x^2/a^2)$
with a parabolic transverse confining potential.
Thus $\epsilon_n^0(x)=\epsilon_0^0(0)\,(2n+1)\,(1+x^2/a^2)^{-2}$,
with the zeroth transverse energy maximum $\epsilon_0^0(0)=\hbar^2/2md^2$.
The generation rate is shown in units of
$m W_{\bf q} \eta^2 a / {\cal V} \hbar^2 q$,
the bias voltage in units of $\epsilon_0^0(0)/e$.
The screening factor $\eta$ is set constant for simplicity.
The quasi Fermi level $\mu$ is chosen such that
$\mu^{(+)}=\epsilon_0^0(0)$ when $eV=\hbar\omega_{\bf q}$.
Furthermore,
$d=0.08\,\mu{\text{m}}$,
$a=6\,\mu{\text{m}}$,
$m=0.07m_0$,
$s=3\times 10^5\,{\text{cm/s}}$,
$q=6ms/\hbar$, and
$\theta=0$.
}
\label{fig8}
\end{figure}

\begin{figure}
\centerline{\psfig{figure=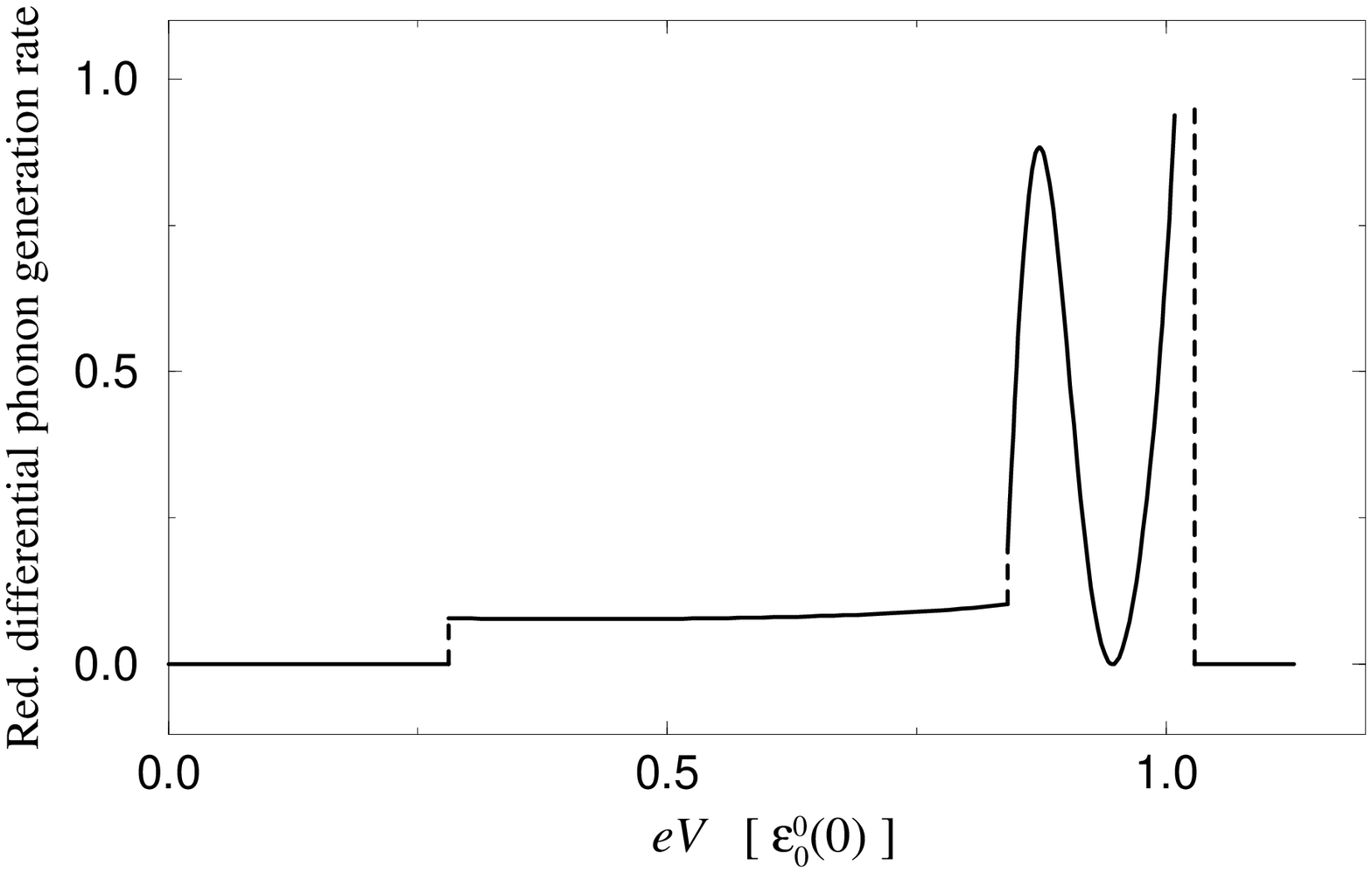,width=12cm}}
\vspace{0,4cm}
\caption{
Dependence of the differential generation rate
$\cal R$
[shown in units of $mW_{\bf q}e\eta^2a/{\cal V}\hbar q\,\epsilon_0^0(0)$]
on the gate voltage $V$ [in units of $\epsilon_0^0(0)/e$].
The differential rate is not continuous at the points
where the transitions change character.
Furthermore, it diverges (in this approximation) for
$\mu^{(+)}\rightarrow k_1^2/2m+\epsilon_0^0(0)$.
The parameters are the same as for Fig.~\ref{fig8}.
}
\label{fig9}
\end{figure}

Figure~\ref{fig8} shows
the dependence of the generation rate on the bias voltage
for a set of typical parameters.
The phonon generation sets on for $eV=\hbar\omega_{\bf q}$ and,
in contrast to the case of a uniform channel,
increases approximately linearly:
As $eV$ increases,
more and more phonon-emitting transitions become possible,
since the energies of the involved electrons is not fixed.
Figure~\ref{fig9} shows
the variation of the differential generation rate with the bias voltage.
It is proportional to the integrand of Eq.~(\ref{conuc2})
with $\epsilon$ replaced by $\mu^{(+)}$.
The slowly varying part corresponds to
transitions from propagating to reflected states,
while the oscillating part corresponds to transitions between
oppositely directed propagating states.

It is important to note that the two patterns of phonon generation
considered in this paper are physically different.
The first one (Sec.~\ref{secuniform}) involves transitions
that take place homogeneously inside the whole channel (quantum wire).
The pattern considered in this section, on the other hand,
involves transitions only near the edges of the channel,
where the conservation laws are local
and thus different from the ones for the uniform part.
As a result, phonons with $\hbar q<2ms$ can be generated.
Observations of such phonons can be
an indication of the importance of the processes near the edges.
An effective interaction length in this case is $l(x^*)=|d(x^*)/d'(x^*)|$.
An estimate of the relative intensity of those processes
compared to uniform generation is given by $\min[l(x^*)/{\cal L},1]$,
where ${\cal L}$ is the approximate channel length.

In this section, we have considered the case of
intraband transitions in a nonuniform channel.
The calculations in the interband case ($n\ne n'$) are similar,
but more complicated, and will not be carried out here.
However, the main effects discussed above, i.e.\ interference
and the possibility to discriminate between different kinds of transitions,
are relevant to that case as well.
Also, as in the case of interband transitions in a uniform channel
(Sec.~\ref{subsecinter}),
there are voltage thresholds in the phonon generation rate.
Due to the $x$ dependence of the band gap, $\Delta_{nn'}(x)$,
these thresholds are less pronounced.

Phonon generation is accompanied by non-Ohmic behavior of the conductance.
This was analyzed for optical phonons in Ref.~\onlinecite{gur2}. For
the case of acoustical phonons it needs both additional experimental
and theoretical investigation.

We have discussed generation of bulk phonons. However, the same methods
can be applied for the investigation of the generation of elastic
modes confined to the interfaces.

\section*{Conclusion}

We have calculated the rate of phonon generation
by a current-carrying quantum channel.
The generation rate is a steplike function of the applied bias voltage.
The threshold voltages are directly related to
the band gaps between the modes of transverse quantization,
while the generation rates at the plateaus are related to
the electron-phonon coupling constant inside the channel.
The emitted phonons have a characteristic angular distribution
with a cutoff at an upper limiting angle.
The rate corresponding to phonons that are generated near the edges
has a characteristic voltage dependence
that is sensitive to the shape of the channel.

In conclusion, we shown that
the spectral and spatial distributions of emitted phonons
as well as
the dependence of the generation rate on the bias voltage
bear information
both on electron-phonon coupling in the quantum channel
and on characteristics of the electron spectrum.

\acknowledgements

V. L. Gurevich is grateful to the University of
Oslo for hospitality and financial support of this work. His work was
also partially supported by the Russian National Fund of Fundamental
Research (Grant No 97-02-18236-a).

H.\ Totland is grateful to the Norwegian Research Council for
financial support.

\newpage  

\vspace{1cm}

\newpage \vspace*{2cm}

\vspace{0,5cm}

\newpage \vspace*{3cm}

\newpage

\vspace{0,6cm}

\begin{references}

\bibitem{KNH} A. J. Kent, A. J. Naylor, P. Hawker, M. Henini, and B. Bracher,
Phys. Rev. B {\bf 55}, 9775 (1997).
\bibitem{TSC} V. I. Talyanskii, J. M. Shilton, J. Cunningham, M. Pepper,
C. J. B. Ford, C. G. Smith, E. H. Linfield, D. A. Ritchie, and G. A. C. Jones,
Physica B {\bf 251}, 140 (1998).
\bibitem{RDC} S. Roshko, W. Dietsche, and L. J. Challis, Phys. Rev. Lett {\bf 80},
3835 (1998).
\bibitem{CWGS} E. Chow, H. P. Wei, S. M. Girvin, and M. Shayegan,
Phys. Rev. Lett {\bf 77}, 1143 (1996).
\bibitem{lby} R. Landauer, IRM J. Res. Dev. {\bf 1}, 233 (1957); {\bf
32}, 306 (1989); Y. Imry, in {\it Directions in Condensed Matter
Physics}, ed. by G. Grinstein and G. Mazenko (World Scientific,
Singapore, 1986), p. 101; M. B\"uttiker, Phys. Rev. Lett. {\bf 57},
1761 (1986).
\bibitem{gur1} V. L. Gurevich, Phys. Rev. B {\bf 55}, 4522 (1997).
\bibitem{end0} I. O. Kulik, R. I. Shekhter, and A. N. Omelyanchouk
[Sol. St. Comm. {\bf 23}, 301 (1977)] pointed out that the processes
leading to electric resistance and heat generation
are spatially separated in a classical point contact.
\bibitem{NAW} N. Nishiguchi, Y. Ando, M. N. Wybourne,
J. Phys.: Condens. Matter {\bf 9}, 5751 (1997).
\bibitem{gur}
V. L. Gurevich, {\em Transport in Phonon Systems} (North-Holland,
Amsterdam, 1986). 
\bibitem{GPI}V. L. Gurevich, V. B. Pevzner, and G. J. Iafrate, Phys. Rev.
Lett. {\bf 77}, 3881 (1996).
\bibitem{gur2} V. L. Gurevich, V. B. Pevzner, and G. J. Iafrate,
Phys. Rev. Lett. {\bf 75}, 1352 (1995); J. Phys.: Condens. Matter {\bf
7}, L445 (1995).
\bibitem{BTG} {\O}.\ L.\ B{\o}, H.\ Totland, and Yu.\ Galperin,
  J. Phys.:\ Condens.\ Matter {\bf 9}, 8381 (1997).
\bibitem{Les}L. I. Glazman, G. B. Lesovik, D. E. Khmel'nitskii, and
 R. I. Shekhter, Pis'ma Zh. Eksp. Teoret. Fiz. {\bf
 48}, 218 (1988) [Sov. Phys. --- JETP Lett. {\bf 48}, 238 (1988)].
\bibitem{end1} We assume the amount of {\em partially} reflected states
is small.
\bibitem{wix} A. Wixworth, J. Schriba, M. Wassermeier, J. P. Kotthaus,
G. Weinmann, and W. Schlapp, Phys. Rev. B {\bf 40}, 7874 (1989).
\bibitem{GGKS} A.~Grincwajg, L.~Y.~Gorelik, V.~Z.~Kleiner, and R.~I.~Shekhter,
  Phys.\ Rev.\ B {\bf 52}, 12168 (1995).
\bibitem{TBG} H.\ Totland, {\O}.\ L.\ B{\o}, and Y.\ M.\ Galperin,
  Phys.\ Rev.\ B {\bf 56}, 15299 (1997).
\bibitem{MG} F.~A.~Maa{\o} and Yu.~M.~Galperin, Phys. Rev. B {\bf 56},
  4028 (1997).
\end{references}
\end{document}